\documentclass[11pt]{article}%
\usepackage{amssymb}
\usepackage{amsfonts}
\usepackage{amsmath}
\usepackage{graphicx}%
\usepackage{subcaption}
\usepackage{harvard}
\usepackage{authblk}
\providecommand{\U}[1]{\protect\rule{.1in}{.1in}}
\usepackage{color}

\citationmode{abbr}
\citationstyle{dcu}
\begin{document}
\title{\textbf{QUANTIFYING THE ROBUSTNESS OF METRO NETWORKS}}
\author[1]{Xiangrong Wang}
\author[2]{Yakup Ko\c c}
\author[3]{Sybil Derrible}
\author[3]{Sk Nasir Ahmad}
\author[1,4]{Robert E. Kooij}
\affil[1]{Faculty of Electrical Engineering, Mathematics and Computer
Science, Delft University of Technology, The Netherlands}
\affil[2]{Systems Engineering Section, Faculty of Technology, Policy and Management, Delft University of Technology, The Netherlands}
\affil[3]{Civil and Materials Engineering Department, University of Illinois at Chicago, U.S.A.}
\affil[4]{TNO (Netherlands Organisation for Applied Scientific Research), Delft, The Netherlands}
\date{}
\maketitle
\begin{abstract}
Metros (heavy rail transit systems) are integral parts of urban transportation systems. Failures in their operations can have serious impacts on urban mobility, and measuring their robustness is therefore critical. Moreover, as physical networks, metros can be viewed as topological entities, and as such they possess measurable network properties. In this paper, by using network science and graph theoretical concepts, we investigate both theoretical and experimental robustness metrics (i.e., the robustness indicator, the effective graph conductance, and the critical thresholds) and their performance in quantifying the robustness of metro networks under random failures or targeted attacks. We find that the theoretical metrics quantify different aspects of the robustness of metro networks. In particular, the robustness indicator captures the number of alternative paths and the effective graph conductance focuses on the length of each path. Moreover, the high positive correlation between the theoretical metrics and experimental metrics and the negative correlation within the theoretical metrics provide significant insights for planners to design more robust system while accommodating for transit specificities (e.g., alternative paths, fast transferring).
\end{abstract}
\newpage
\section{INTRODUCTION}

With constant urbanization  \cite{united_nations_world_2014}, cities around the world are not only growing in number but they are also growing in size. As one of the main modes of urban transportation, public transit systems are integral to move people efficiently in cities \cite{vuchic_urban_2005}. Indeed, they provide myriads of benefits, from reducing traffic congestion to having a lesser impact on the environment, emitting fewer greenhouse-gases per capita than the conventional automobile. The future of public transportation is therefore bright. While increasing transit use is desirable, effort must be put into developing designs that are also resilient and robust. These subjects have gathered much interest in the scientific community in recent years, especially within the context of resilience to extreme events \cite{Risk2014,cai2014increasing}. Resilience typically refers to the ability to return to a previous state after a disruption, while robustness tends to measure the amount of stress that can be absorbed before failure.

%Traditionally, transit resilience and robustness have been associated largely with travel time reliability and variability \cite{levinson2005reliability}. It is still an important topic today from the quantifying variability itself \cite{Kieu2015,Mazloumi2010623} or its cost \cite{Benezech20131}, to using reliability and variability as design criteria \cite{Yao2014233,an2014service}.

Traditionally, transit resilience and robustness have been associated largely with travel time reliability and variability \cite{levinson2005reliability}. It is still an important topic today from quantifying variability itself \cite{Kieu2015,Mazloumi2010623} or its cost \cite{Benezech20131}, to using reliability and variability as a design criterion \cite{Yao2014233,an2014service}. Recently, the field of \emph{Network Science} \cite{newman_networks_2010} has emerged as particularly fitted to measure the robustness of a system, notably by studying the impact of cascading failure \cite{watts_simple_2002,crucitti_model_2004,kinney_modeling_2005}. Indeed, as physical networks, metros are composed of stations (nodes) and rail tracks (links), and they therefore possess measurable network properties \cite{derrible_network_2009,derrible_characterizing_2010} that can be used to study their robustness \cite{berche_resilience_2009,derrible2010complexity,von_ferber_tale_2012}. Several works have also tried to combine information from both transit operation and network properties to gain insight into the robustness of transit networks \cite{rodriguez-nunez_measuring_2014,kim_network_2015}.

In this work, our main objective is to define, use and compare multiple indicators and metrics to quantify the robustness of 33 worldwide metro systems within the realms of graph theory and network science. Metro, here, refers to heavy rail transit systems, whether underground, at grade, or overground.

To assess the robustness of metros, our main research approach is to subject metros to random failures and targeted attacks. Three different metrics are used to compare the response of the metro networks to failures and attacks: (i) the robustness indicator $ r^T $, see \cite{derrible2010complexity}, (ii) the effective graph resistance $ R_G $, see \cite{PVM_graphspectra} and (iii) the critical thresholds $ f_{90\% }$ and $ f_c $, see for instance \cite{cohen2010complex}. The first two metrics are analytical expressions available from the existing complex network literature, the critical thresholds are obtained through simulation.

For this work, data from \cite{derrible_network_2012} was used. In this dataset, only terminals and transfer stations are included as opposed to all stations. The other stations were omitted on purpose since they tend to bias the results by simply connecting with two adjacent stations. For more details on the data, the reader is referred to \cite{derrible2010complexity,derrible_network_2012}.

\section{ROBUSTNESS METRICS}
This section elaborates on the robustness metrics that are calculated in this study. Sequentially, we first introduce the robustness indicator $ r^T $, followed by the effective graph conductance $ C_G $, and finally the critical thresholds $ f_{90\%} $ and $ f_c $.

\subsection{The robustness indicator $ r^T $}

The \emph{robustness indicator $r^T$} is suggested as a robustness metric for metro networks by Derrible and Kennedy \cite{derrible2010complexity}. It quantifies the robustness of a metro network in terms of the number of alternative paths in the network topology divided by the total number of stations in the system:

\begin{equation}
r^T = \frac{\mu - L^m}{N_{S}}
\label{eqRobustnessIndicatorOriginal}
\end{equation}
where $N_{S}$ is the total number of stations (not limited to transfers and terminals), $L^m$ is the number of multiple links between two nodes (e.g., overlapping lines), and $\mu$ is the cyclomatic number that calculates the total number of alternative paths in a graph; $\mu = L - N + P$, with $L$ the number of links, $N$ the number of nodes, and $P$ the number of subgraphs. Transit networks are typically connected and $P=1$. The total number of stations, $N_{S}$ in the denominator represents a likelihood of failure; i.e., the larger the system, the more stations need to be maintained, and therefore the more likely a station may fail.

For this work, we do not consider any multiple edges\footnote{Even when two stations are directly connected by multiple lines, we assign a value of 1 to the adjacency matrix. The definition is given in Section \ref{subsectionMetroNetwork}}. Moreover, we also use the number of nodes $N$ (i.e., transfer stations and terminals) in the denominator as opposed to the total number of stations $N_{S}$. Equation \ref{eqRobustnessIndicatorOriginal} therefore becomes:

\begin{equation}
r^T=\frac{L-N+1}{N}
\label{eqRobustnessIndicator}
\end{equation} 

Essentially, $r^T$ increases when alternative paths are offered to reach a destination, and it decreases in larger systems, which are arguably more difficult to upkeep.

%where $ L $ denotes the number of links and $ N $ is the number of stations in the metro network.

%The definition of the robustness indicator in equation \eqref{eqRobustnessIndicator} indicates that increasing the number of alternative paths is a positive element for the robustness. On the other hand, the robustness indicator characterizes the fact that increasing the number of stations induces a higher probability of random failures. The probability that at least one station fails in the whole metro network is $ \text{Prob[\{at least one station fails\}]}=1-(1-\frac{1}{N})^N $ by assuming that all the stations have an equal probability $ \frac{1}{N} $ to fail. The derivative of the probability $ \text{Prob}^{'}[ \{\text{at least one station fails }\}]=-\frac{(N-1)^{(N-1)}}{N^2} $ is nonpositive and therefore the probability is a decreasing function of the number $ N $ of the stations.

\subsection{The effective graph conductance $ C_G $}
The \emph{effective graph resistance $ R_G $} captures the robustness of a network by incorporating the number of parallel paths (i.e., redundancy) and the length of each path between each pair of nodes. The existence of parallel paths between two nodes in metro networks and a heterogeneous distribution of each path length result in a smaller effective graph resistance and potentially a higher robustness level. 

The effective resistance $ R_{ij} $ \cite{PVM_graphspectra} between a pair of nodes $ i $ and $ j $ is the potential difference between these nodes when a unit current is injected at node $ i $ and withdrawn at node $ j $. The effective graph resistance $ R_G $ is the sum of $ R_{ij} $ over all pairs of nodes in the network. An efficient method for the computation of the effective graph resistance in terms of the eigenvalues is 

\begin{equation}
R_G=N\sum_{i=1}^{N-1}\frac{1}{\mu_i}
\end{equation}
where $ \mu_i $ is the $ i $th non-zero eigenvalue of the Laplacian matrix\footnote{An $ N \times N $ matrix representing the graph. The definition is given in Section \ref{subsectionMetroNetwork}.}. Properties of the effective graph resistance are given in \cite{PVM_graphspectra}. The effective graph resistance is considered as a robustness metric for complex networks \cite{wang2014improving}, especially for power grids \cite{kocc2014impact,kocc2014structural}. In this paper, we use a normalized version of the effective graph resistance, called the \emph{effective graph conductance}, defined as 

\begin{equation}
C_G=\frac{N-1}{R}
\label{eqEffGraphConducatance}
\end{equation}
where $ C_G $ satisfies $ 0 \leq C_G \leq 1 $. Here, a larger $ C_G $ indicates a higher level of robustness.

\subsection{Critical thresholds}

\emph{Critical thresholds} relate to the fraction of nodes that have to be removed from the network, such that the size of the largest connected component of the remaining network is equal to a predetermined fraction of the size of the original network. Critical thresholds, which are also used in the percolation model \cite{newman2002percolation,callaway2000network}, characterize the robustness of interconnection patterns with respect to the removal/failure of network nodes.

In this paper, we first consider the threshold $ f_{90\%} $, the fraction of nodes that have to be removed such that the remaining network has a largest connected component that contains $ 90\% $ of the original network. For the node removal process, we simulate both random failures and targeted attacks. In the case of \emph{random failures}, the nodes are removed by random selection, while for \emph{targeted attacks}, the nodes are removed progressively based on their degrees (i.e., stations with many connections are removed first).

For the targeted attack, we also consider the \emph{critical threshold $ f_c $} defined as the fraction of nodes to be removed such that the largest component is reduced to a size of one node (i.e., the network is completely disintegrated). Differently than the first two metrics, the critical thresholds $ f_{90\%} $ and $ f_c $ are obtained through simulations.

%The simulation results of the critical thresholds are considered as the ground-truth and the performance of the robustness indicator and the effective graph conductance are evaluated by comparing them with the critical thresholds. 

\section{EXPERIMENTAL METHODOLOGY}

The experimental method considers 33 metro networks and subjects them to failures or to deliberate attacks to determine their robustness. This approach can be used to evaluate the performance of different robustness metrics for metro networks under node failures/attacks. This section elaborates on the metro networks, attack strategies and the evaluation of robustness for metro networks.

\subsection{Metro Networks}
\label{subsectionMetroNetwork}
We define metros as urban rail transit systems with exclusive right-of-way whether they are underground, at grade or elevated. The physical metro network can be represented by an undirected graph $ G(N,L) $ consisting of $ N $ nodes and $ L $ links. As mentioned, the nodes are transfer stations and terminals, while the links are rail tracks that physically join stations. A graph $ G $ can be completely represented by an adjacency matrix $ A $ that is an $ N \times N $ symmetric matrix with element $ a_{ij}=1 $ if there is a line between nodes $ i $ and $ j $, otherwise $ a_{ij}=0 $. The Laplacian matrix $Q=\Delta-A$ of $G$ is an $ N \times N $ matrix, where $ \Delta=\text{diag}(d_i) $ is the $ N \times N $ diagonal degree matrix with the elements $ d_i=\sum_{j=1}^{N}a_{ij} $. The eigenvalues of $Q$ are non-negative and at least one is zero \cite{PVM_graphspectra}. The eigenvalues of $Q$ are ordered as $%
0=\mu _{N}\leq \mu _{N-1}\leq \ldots \leq \mu _{1}$. The degree $ d_i=\sum_{j=1}^{N}a_{ij} $ of a node $ i $ is the number of connections to that node. The degree for the terminals is one.

\begin{figure}[!htp]\centering

\begin{subfigure}[b]{0.45\columnwidth}
\includegraphics[width=\textwidth]{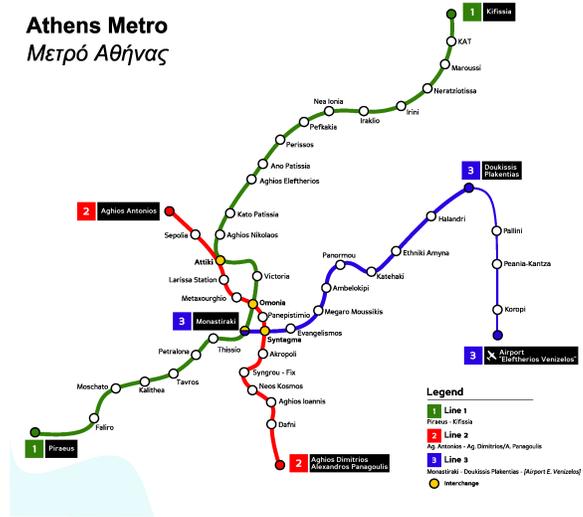}
\caption{The map of the Athens metro network.}
\end{subfigure}

\begin{subfigure}[b]{0.45\columnwidth}
\includegraphics[width=\textwidth]{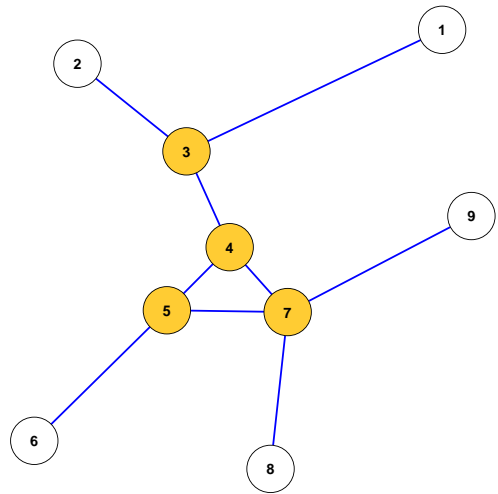}
\caption{The graphical representation of the Athens metro network}
\end{subfigure}
\caption{Athens metro network}

\label{TopologyMetroNetwork}

\end{figure}

In this paper, we look at 33 metro networks. Figure \ref{TopologyMetroNetwork}a shows the map of the Athens metro network\footnote{Adapted from http://commons.wikimedia.org/wiki/File:Athens\_Metro.svg} and the graphical representation is shown in Figure \ref{TopologyMetroNetwork}b. In Figure \ref{TopologyMetroNetwork}b, stations 1 to 9 are respectively: Kifissia, Aghios Antonios, Attiki, Omonia, Monastiraki, Pireaus,  Syntagma, Aghios Dimitrios, and Airport Eleftherios Venizelos. For more details on the methodology, see \cite{derrible2010complexity}.

\subsection{Attack strategies}
To determine the robustness of metro networks, the response of metro networks to targeted attacks or random failures is investigated. This paper considers two strategies for node removal: (i) random node removal and (ii) degree-based node removal. 

\begin{itemize}
\item \emph{Random removal:} The node to be removed is chosen at random from all the nodes in the network with equal probability. 
\item \emph{Degree-based removal:} The node to be removed has the highest degree in the network. If multiple nodes have the highest degree, one node is chosen at random from all the highest-degree nodes with equal probability. In this paper, nodes are removed progressively. We first remove the node with highest degree, and continue selecting and removing nodes in decreasing order of their degree.
\end{itemize}
%\subsection{Robustness evaluation}

%As physical networks, metros are composed of stations (nodes) and rail tracks (links), and they therefore possess measurable network properties, such as degree distribution and clustering coefficient. The robustness of metro systems impacts their ability to offer alternative paths to transit users during the occurrence of failures, accidents or even targeted attacks. We propose robustness metrics that capture the impact of the \emph{topology of transit network} on the robustness of metro networks under targeted attacks. 

\begin{figure}[!htp]\centering
\begin{subfigure}[b]{0.44\textwidth}
\includegraphics[width=\textwidth]{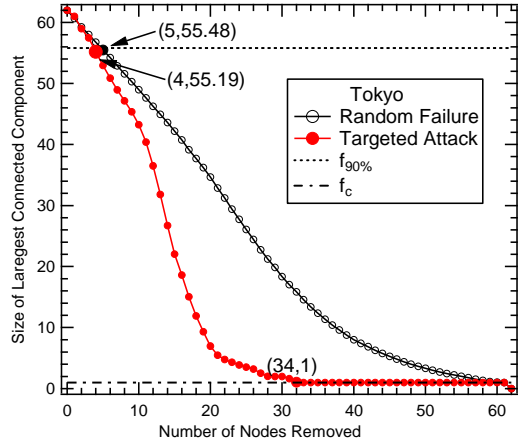}
\end{subfigure}
\caption{The robustness curve for the Tokyo metro network}
\label{ExampleRobustnessCurve}
\end{figure}

After a node is removed, the size of the largest connected component of the remaining network is determined. Measuring the size of the largest connected component for an interval of removed nodes $ [1,N] $ results in a robustness curve. According to the robustness curve, we then determine the critical thresholds $ f_{90\%} $ and $ f_c $. The critical threshold $ f_{90\%} $ is the first point at which the size of the largest connected component is less than $ 90\% $ of the original network size. When determining the $ f_{90\%} $ for random node removal, the size of the largest connected component is the average of $ 1000 $ simulation runs. Similarly, the critical threshold $ f_c $ is the first point at which the size of the largest connected component is one (i.e., the network is completely disintegrated). Figure \ref{ExampleRobustnessCurve} exemplifies the determination of the critical thresholds from the robustness curve in \emph{Tokyo} metro network with $ 62 $ nodes. Computing the size of the largest connected component for removed nodes from $ 1 $ to $ 62 $ results in a robustness curve. The size of the largest connected component is $ 56.77 $ after randomly removing $ 4 $ nodes. After removing $ 5 $ nodes, the size becomes $ 55.48 $ which is smaller than $ 90\% \times 62=55.8 $, i.e., $ 90\% $ of the size of the network. Therefore, the critical threshold $ f_{90\%} $ is determined as $ \frac{5}{62} $. The threshold $ f_c $ is determined in a similar way. The critical thresholds are regarded as the experimental robustness level of metro networks with respect to node failures. 

\section{NUMERICAL ANALYSIS}

In this section, we calculate and compare the results for the three measures discussed above. Firstly, the robustness metrics $ r^T $ and $ C_G $ are computed for the 33 metro networks. Secondly, the critical thresholds of metro networks under random failures and targeted attacks are determined by simulations. Finally, the relationship between the three measures are compared and correlated.

\subsection{Effectiveness of robustness metrics}

Table \ref{RobustnessMetricsMetroNetwork} shows the values of the robustness metrics $ r^T $ (column 3) and $ C_G $ (column 4) computed by equations \eqref{eqRobustnessIndicator} and \eqref{eqEffGraphConducatance} for the 33 metro networks.

According to the rank of the robustness indicator $ r^T $, the most robust network is \emph{Tokyo} with $ r^T=0.74 $, followed by \emph{Madrid } and \emph{Paris} with $  r^T=0.67$ and $ 0.62 $, respectively. Moreover, \emph{Seoul}, \emph{Moscow} and \emph{MexicoCity} also have a relatively high robustness level. Clearly, the robustness indicator $r^T$ favors larger networks that have developed many alternative paths between any pairs of nodes. At the same time, $r^T$ discredits networks that have a high number of nodes while having few alternative paths. This is particularly exemplified by the case of \emph{New York}. Due to the topography of the region, the \emph{New York} metro lines run mostly North-South from the Bronx to Lower Manhattan and East-West in Queens and Brooklyn. The lines therefore seldom intersect as opposed to the case of the \emph{Seoul} metro for instance.

\begin{figure}[!htp]\centering
\begin{subfigure}[b]{0.24\textwidth}
\includegraphics[width=\textwidth]{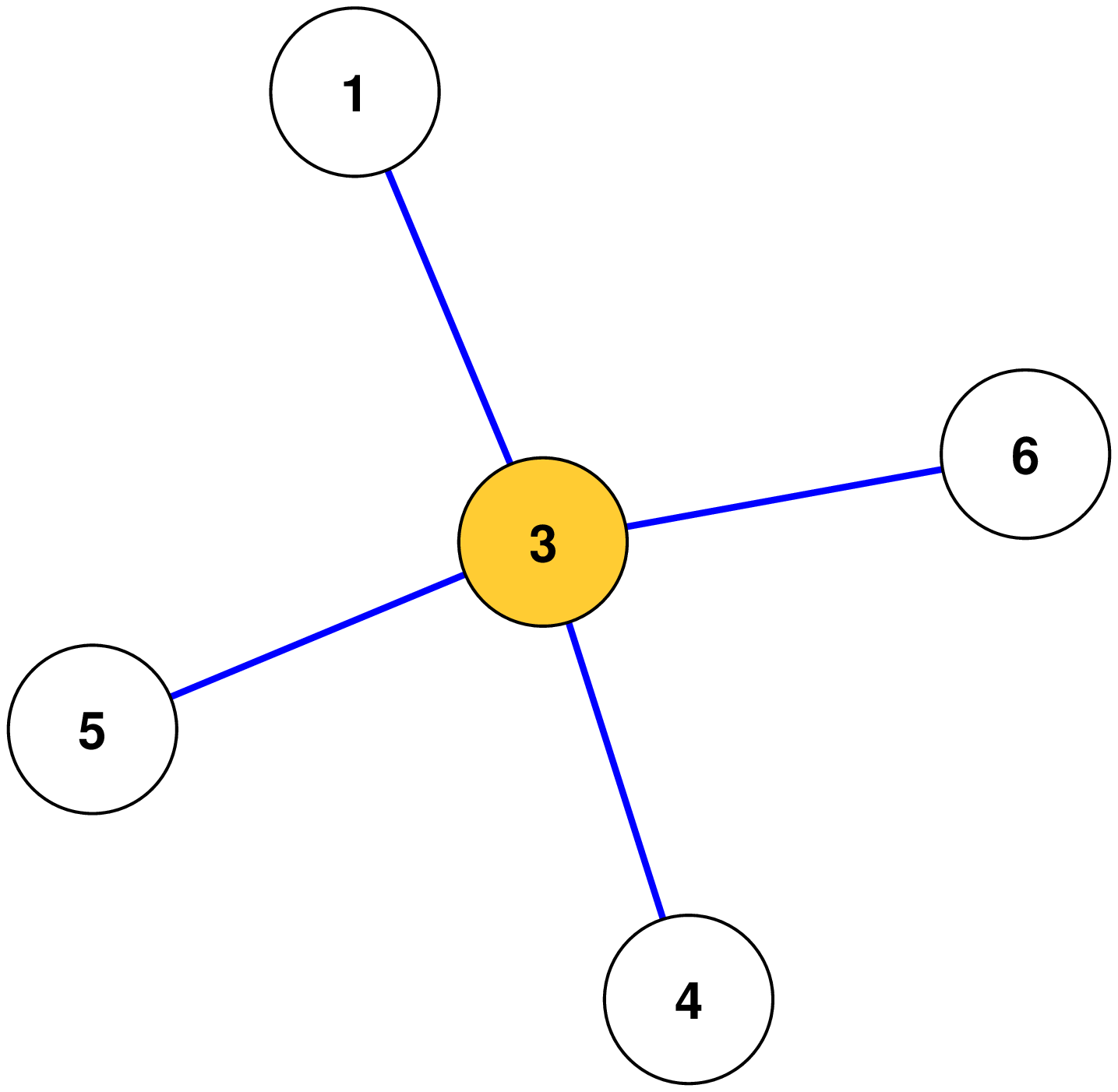}
\caption{Rome}
\end{subfigure}
\begin{subfigure}[b]{0.24\textwidth}
\includegraphics[width=\textwidth]{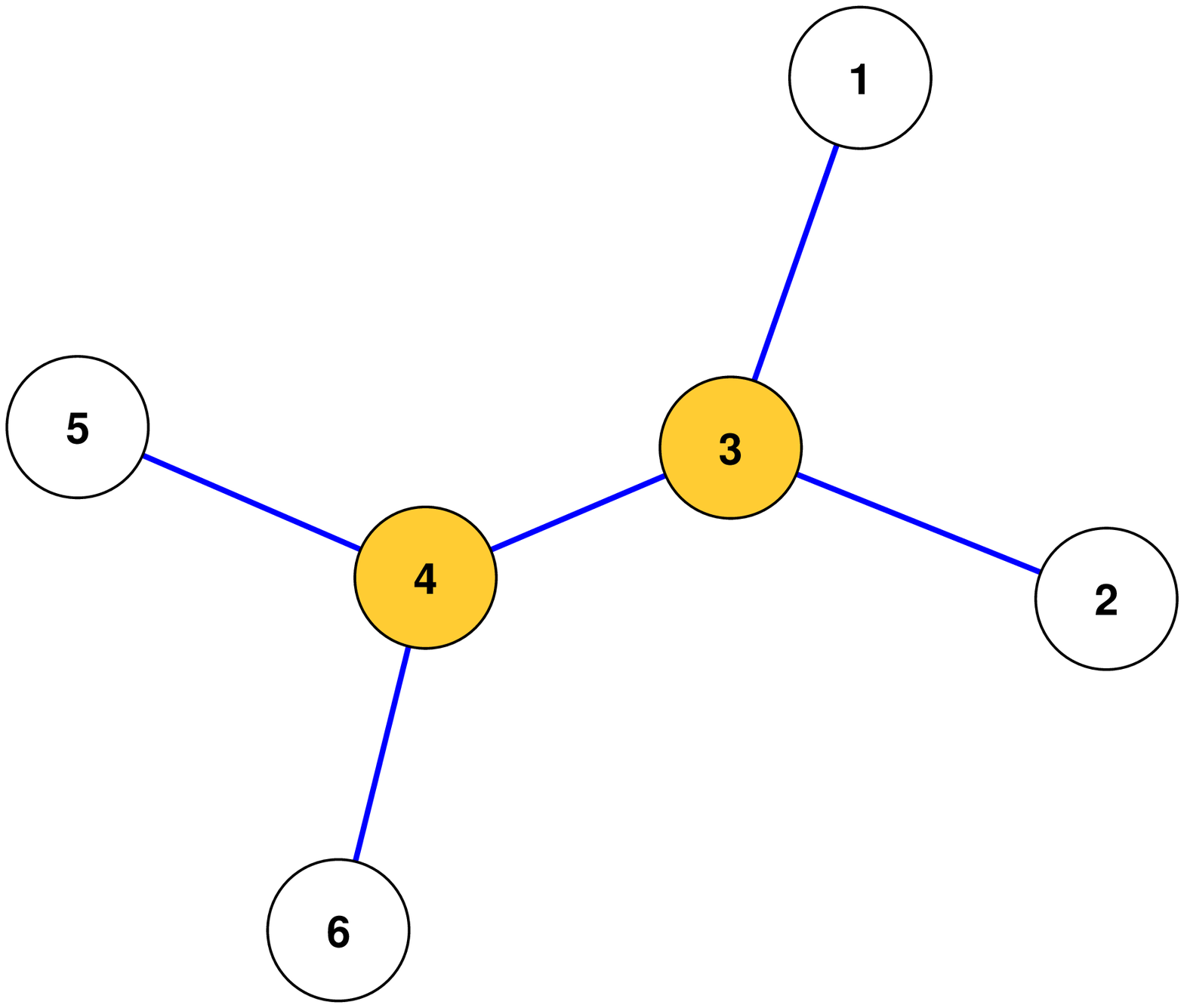}
\caption{Cairo and Marseille}
\end{subfigure}
\begin{subfigure}[b]{0.24\textwidth}
\includegraphics[width=\textwidth]{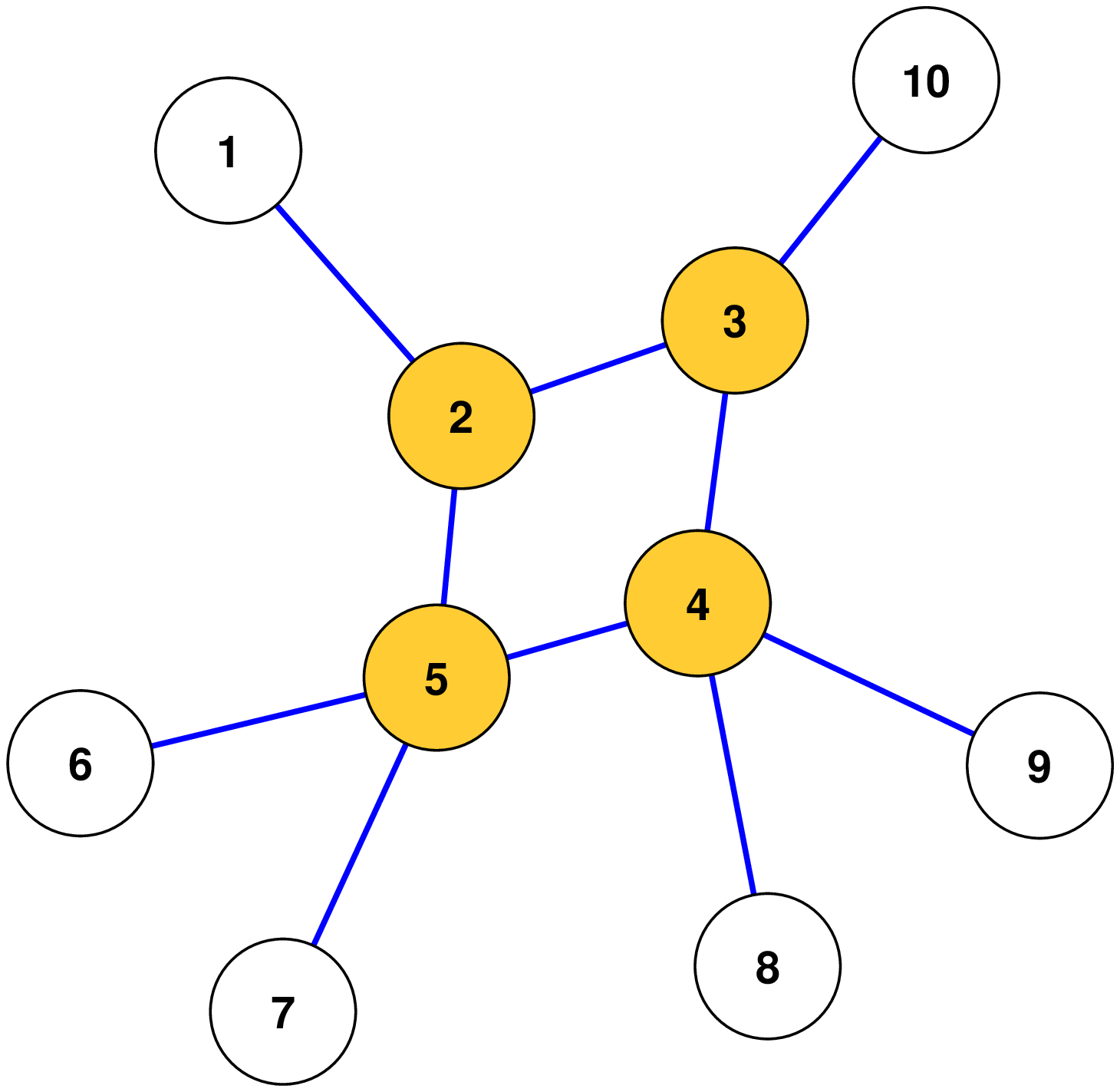}
\caption{Montreal}
\end{subfigure}
\begin{subfigure}[b]{0.24\textwidth}
\includegraphics[width=\textwidth]{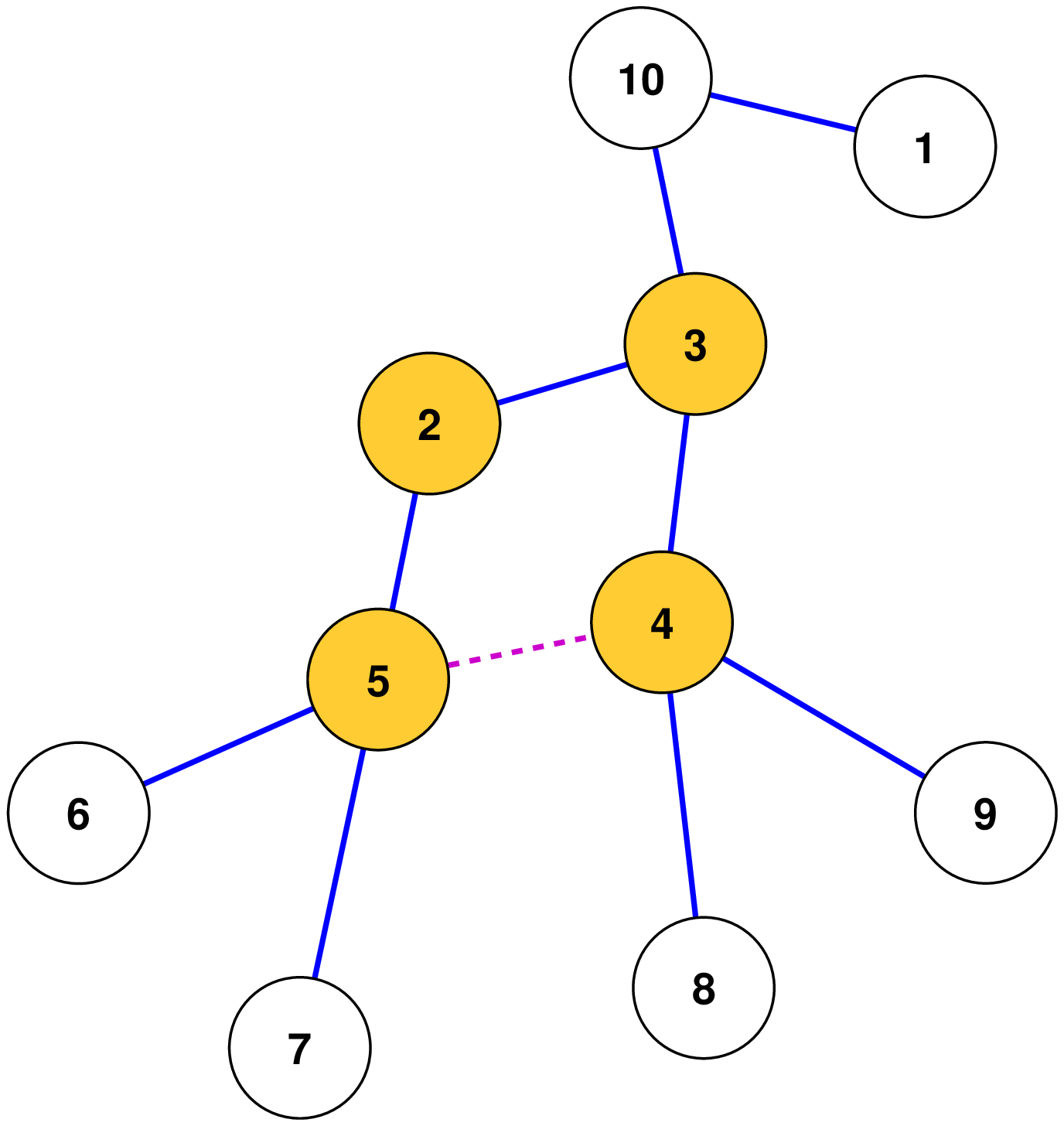}
\caption{Toronto}
\end{subfigure}
\caption{The topology of metro networks}
\label{TopologyofMetroNetworks}
\end{figure}

According to the effective graph conductance $ C_G $, \emph{Rome} with $ C_G=0.25  $ has the highest robustness level, followed by \emph{Cairo} and \emph{Marseille} both with $ C_G=0.17 $. The effective graph conductance not only accounts for the number of alternative paths, but it also considers the length of each alternative path. For smaller networks without cycles (e.g., star graph), the effective graph conductance increases due to the lower average path length between two stations. The topologies in Figure \ref{TopologyofMetroNetworks}a and Figure \ref{TopologyofMetroNetworks}b are particular examples. In this case, a higher effective graph conductance indicates a faster transfer between two transit stations. At the same time, effective graph conductance favors networks with the smallest length of the shortest paths. Taking Figure \ref{TopologyofMetroNetworks}c (\emph{Montreal}) and Figure \ref{TopologyofMetroNetworks}d\footnote{In order to compare the topology of \emph{Montreal} and \emph{Toronto}, a link between stations $ 4 $ and $ 5 $ is added into \emph{Toronto} and the effective graph conductance is $ 0.099 $.} (\emph{Toronto}) as examples, the difference between the topologies is that station $ 1 $ connects to $ 10 $ and then connects to station $ 3 $ in \emph{Toronto}, while stations $ 1 $ and $ 10 $ separately connect to stations $ 2 $ and $ 3 $ in \emph{Montreal}. The total length of shortest paths from station $ 1 $ to the rest of the stations is higher in \emph{Toronto} than in \emph{Montreal}. Compared to \emph{Toronto}, the higher effective graph conductance in \emph{Montreal} indicates that the effective graph conductance favors the star-like topology with a smaller average shortest path length.

Both analytical robustness metrics, $ r^T $ and $ C_G $, therefore capture different aspects of metro network design.

\begin{table}[!htp]
\caption{Robustness metrics in 33 metro networks}
  \centering
    \begin{tabular}{cccccccc}
    \hline
    \textbf{Networks} & \textbf{N} & \textbf{L} & $ r^T $& \textbf{$ C_G $} & \textbf{$ f_{90\%} $-Degree} & \textbf{$ f_{90\%} $-Random} & \textbf{$ f_c $} \\ \hline
Athens  & 9  & 9  & 0.11  & 0.11  & 0.11  & 0.11  & 0.33 \\
Barcelona  & 29  & 42  & 0.48  & 0.03  & 0.07  & 0.03  & 0.55\\
Berlin  & 32  & 43  & 0.38  & 0.03  & 0.09  & 0.06  & 0.53\\
Boston  & 21  & 22  & 0.1  & 0.03  & 0.05  & 0.05  & 0.38\\
Brussels  & 9  & 9  & 0.11  & 0.11  & 0.11  & 0.11  & 0.44\\
Bucharest  & 11  & 12  & 0.18  & 0.1  & 0.09  & 0.09  & 0.45\\
BuenosAires  & 12  & 13  & 0.17  & 0.09  & 0.08  & 0.08  & 0.33\\
Cairo  & 6  & 5  & 0  & 0.17  & 0.17  & 0.17  & 0.33\\
Chicago  & 25  & 30  & 0.24  & 0.03  & 0.08  & 0.04  & 0.44\\
Delhi  & 8  & 7  & 0  & 0.12  & 0.13  & 0.13  & 0.25\\
HongKong  & 17  & 18  & 0.12  & 0.04  & 0.06  & 0.06  & 0.47\\
Lisbon  & 11  & 11  & 0.09  & 0.09  & 0.09  & 0.09  & 0.36\\
London  & 83  & 121  & 0.47  & 0.01  & 0.07  & 0.06  & 0.51\\
Lyon  & 10  & 10  & 0.1  & 0.11  & 0.10  & 0.10  & 0.4\\
Madrid  & 48  & 79  & 0.67  & 0.03  & 0.08  & 0.10  & 0.56\\
Marseille  & 6  & 5  & 0  & 0.17  & 0.17  & 0.17  & 0.33\\
MexicoCity  & 35  & 52  & 0.51  & 0.03  & 0.09  & 0.06  & 0.57\\
Milan  & 14  & 15  & 0.14  & 0.06  & 0.07  & 0.07  & 0.43\\
Montreal  & 10  & 10  & 0.1  & 0.11  & 0.10  & 0.10  & 0.4\\
Moscow  & 41  & 62  & 0.54  & 0.03  & 0.07  & 0.10  & 0.54\\
NewYork  & 77  & 109  & 0.43  & 0.01  & 0.06  & 0.04  & 0.52\\
Osaka  & 36  & 51  & 0.44  & 0.03  & 0.08  & 0.06  & 0.53\\
Paris  & 78  & 125  & 0.62  & 0.01  & 0.08  & 0.06  & 0.53\\
Prague  & 9  & 9  & 0.11  & 0.12  & 0.11  & 0.11  & 0.33\\
Rome  & 5  & 4  & 0  & 0.25  & 0.20  & 0.20  & 0.2\\
Seoul  & 71  & 111  & 0.58  & 0.01  & 0.08  & 0.08  & 0.59\\
Shanghai  & 22  & 28  & 0.32  & 0.04  & 0.09  & 0.09  & 0.45\\
Singapore  & 12  & 13  & 0.17  & 0.08  & 0.08  & 0.08  & 0.5\\
StPetersburg  & 14  & 16  & 0.21  & 0.07  & 0.07  & 0.07  & 0.43\\
Stockholm  & 20  & 19  & 0  & 0.02  & 0.05  & 0.05  & 0.4\\
Tokyo  & 62  & 107  & 0.74  & 0.02  & 0.08  & 0.06  & 0.55\\
Toronto  & 10  & 9  & 0  & 0.07  & 0.10  & 0.10  & 0.4\\
WashingtonDC  & 17  & 18  & 0.12  & 0.04  & 0.06  & 0.06  & 0.35\\
    \hline
    \end{tabular}%
\label{RobustnessMetricsMetroNetwork}%
\end{table}%

\begin{figure}[!htp]\centering
\begin{subfigure}[b]{0.44\textwidth}
\includegraphics[width=\textwidth]{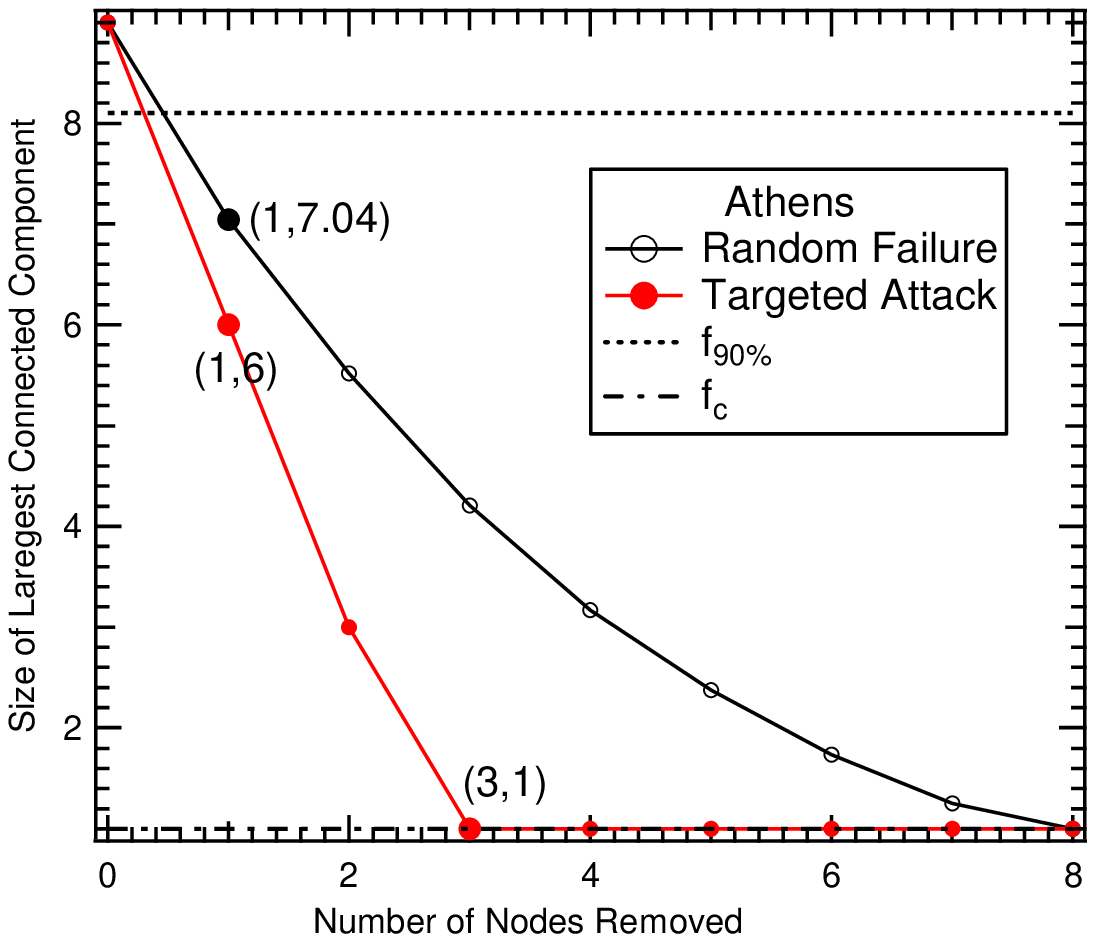}
\caption{Athens}
\end{subfigure}
\begin{subfigure}[b]{0.44\textwidth}
\includegraphics[width=\textwidth]{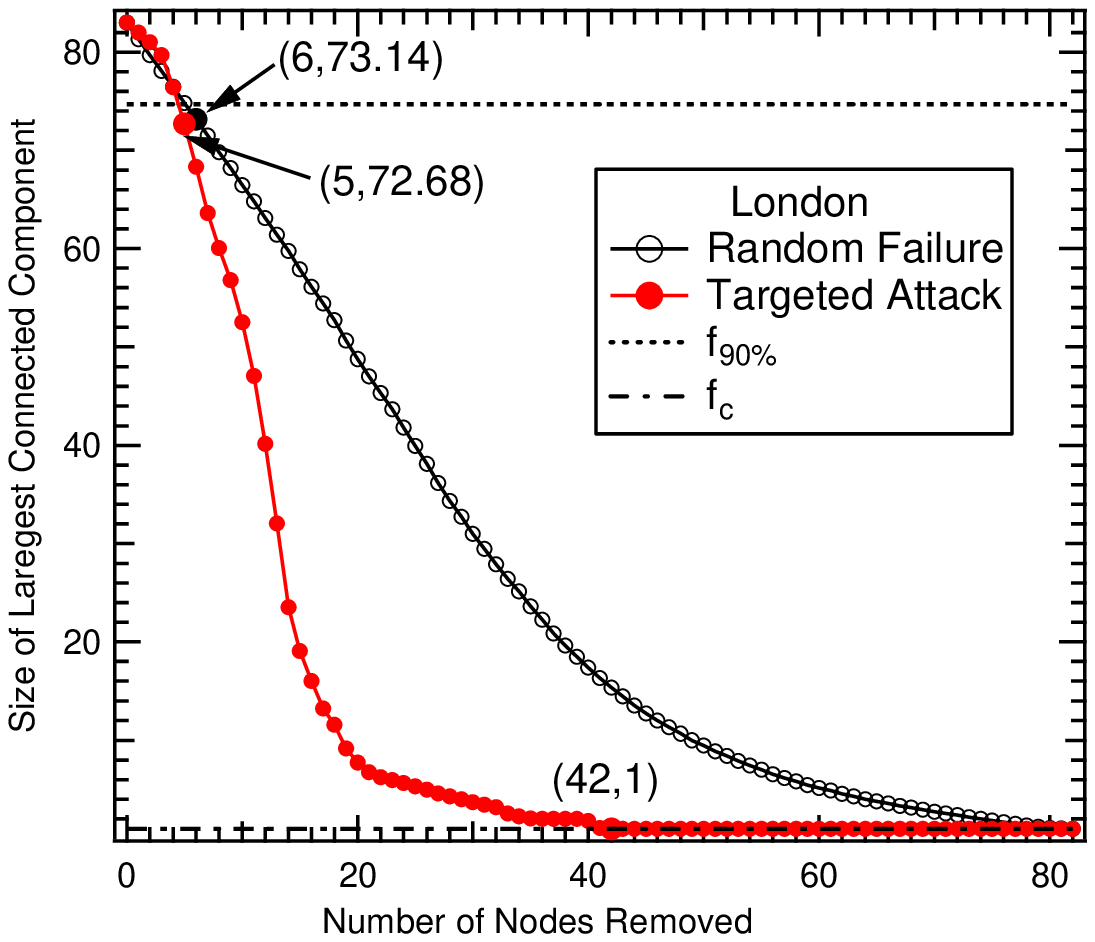}
\caption{London}
\end{subfigure}
\caption{Critical thresholds in metro networks under nodes removal}
\label{CriticalThresholdsMetroNetworks}
\end{figure}

Studying critical thresholds, Figure \ref{CriticalThresholdsMetroNetworks} shows the robustness level of metro networks, taking the \emph{Athens} and \emph{London} metro networks as examples, under random failures and deliberate attacks. The corresponding critical thresholds $ f_{90\%} $ for targeted attacks (column 6) and random failures (column 7), and $ f_c $ for targeted attacks (column 8) are shown in Table \ref{RobustnessMetricsMetroNetwork}. Column 7 and 8 in Table \ref{RobustnessMetricsMetroNetwork} show similar behavior of $ f_{90\%} $ for targeted attacks and random failures.

Similar to the effective graph conductance $ C_G $, \emph{Rome} has the highest robustness level with $ f_{90\%}=0.20 $ both for targeted attacks and random failures. \emph{Cairo} and \emph{Marseille} have the second highest robustness level with $ f_{90\%}=0.17 $ for both targeted attacks and random failures. On the other hand, and similar to the robustness indicator $r^T$, an evaluation of the critical threshold $ f_c $ shows that \emph{Seoul} is the most robust network. It has a critical threshold $ f_c=0.59  $ indicating that $ 59\% $ of nodes need to be removed before the network collapses. \emph{MexicoCity}, \emph{Madrid}, \emph{Barcelona} and \emph{Tokyo} are among the top $ 5 $ of the most robust networks.

To further compare and correlate the three metrics, Table \ref{CorrelationRobustnessMetricCriticalThreshold} shows the Pearson correlation $ \rho $ between the two robustness metrics and the critical thresholds in the metro networks. The high correlation between $ C_G $ and $ f_{90\%} $ for random failures and targeted attacks shows that $ C_G $ effectively captures the $ 10\% $ failure of the metro networks under node removal. The high correlation between $ r^T $ and $ f_c $ shows that $ r^T $ effectively characterizes when the network collapses under node removal.

One explanation for the high correlation between $ r^T $ and $ f_c $ is that the robustness indicator $ r^T $ and the critical threshold $ f_c $ both characterize the number of alternative paths. However, $ r^T $ and $ f_c $ can not be completely correlated (i.e. $ \rho=1 $) due to the scale-free properties in metro networks as shown in \cite{derrible2010complexity}. $ C_G $ and $ r^T $ therefore capture different aspects of metro networks, both of which are important for robustness. However, we should note that the correlation between $ C_G $ and $ r^T $ is $ -0.68 $.

These findings suggests that maximizing one of the two metrics likely decreases the other metric. This is therefore a major issue, which is not atypical of any robustness study. Indeed, while it is easy to develop design recommendations that can make a system more robust to certain conditions, it is much more challenging to develop recommendations that can make a system more robust overall. This point emphasizes the need to use multiple criteria when assessing the design of metro networks. It also points to the fact that robustness (and \emph{resilience} more generally) are terms that are difficult to define and that cannot be solved with a simple objective function within an operations research context \cite{taleb_antifragile_2012}. Instead, much work remains to be done to successfully come up with clear guidelines to transit planners, and simulation and network science may play an important part towards that end.

\begin{table}[!htp]
\centering
\caption{Pearson correlation $ \rho $ between robustness metrics $ r^T $, $ C_G $ and the critical thresholds.}
\begin{tabular}{cccc}
\hline
& \textbf{$ f_{90\%} $-Degree} & \textbf{$ f_{90\%} $-Random} & $ f_c $ \\ \hline
    $ C_G $  & 0.89 & 0.89 & -0.79 \\
    $ r^T $    & -0.40 & -0.45 & 0.85 \\ \hline
\end{tabular}%
\label{CorrelationRobustnessMetricCriticalThreshold}%
\end{table}

\section{CONCLUSION}

The main objective of this work was to investigate the robustness of metro networks by defining and comparing three measures: (1) robustness indicator $r^T$, (2) effective graph conductance $C_G$, and (3) critical thresholds $f$. The two first measures are analytical, while the last one is simulation-based. Moreover, for the critical thresholds $f$, we investigated cases when 90\% of the network was still remaining, $f_{90\%}$ (both under random failure and targeted attack), and when the complete network was disintegrated, $f_c$ (under targeted attack).

Overall, we find that both $r^T$ and $C_G$ capture different aspects of the robustness of metro networks. The former focuses on the number of alternative paths, therefore favoring large networks, and the latter focuses on the length of the paths, therefore favoring small networks. Moreover, we found that the results for $r^T$ aligned closely with $f_c$, while the results for $C_G$ aligned closely with $f_{90\%}$. These results are somewhat problematic since the robustness indicators contradict each other. This finding simply points out the difficulty to account for and measure robustness in a holistic fashion, which therefore calls for more work on this topic so as to eventually help planners design more robust transit systems in this century of cities.

\bibliographystyle{dcu}
\bibliography{Bib/biblio}
%\appendix

%\section{Appendix}
%The definition of robustness indicator $ r^T $ is \cite{derrible2010complexity}
%\begin{equation*}
%r^T=\frac{L-N+1}{N}
%\end{equation*}
%Defining $ L+1 = f(N) $, we have
%\begin{equation*}
%r^T=\frac{f(N)}{N}-1
%\end{equation*}
%In order to make sure that $ r^T $ is an increasing function of $ N $, the derivative of $ r^T $ should be positive:
%\begin{equation*}
%(r^T) '=\frac{f'(N)N-f(N)}{N^2}>0
%\end{equation*}
%from which
%\begin{equation}
%Nf'(N)-f(N)>0
%\label{ConditionforL}
%\end{equation}
%One solution for \eqref{ConditionforL} is $ f(N)=aN^2+bN$, where $ a $ and $ b $ are constant. Meanwhile, the number of links $ L $ in a connected graph is lower bounded by $ N-1 $ and upper bounded by $ \binom{N}{2} $. Therefore, when the network increases, the number of links $ L $ should satisfy $ L=aN^2+bN $,  where $ 0 \leq  a \leq \frac{1}{2}$, b are constant, in order to increase $ r^T $.

\end{document}